# Controlling a remotely located Robot using Hand Gestures in Real-time: A DSP implementation

Jagdish Lal Raheja, Gadula A.Rajsekhar

Machine Vision Lab
CEERI-CSIR
Pilani, Rajasthan, India - 333031
jagdish@ceeri.ernet.in, garaj32@gmail.com

Ankit Chaudhary

Dept. of Computer Science
Northwest Missouri State University
Maryville, MO USA 64468
dr.ankit@ieee.org

**Abstract**–Telepresence is a necessity for present time as we can't reach everywhere and also it is useful in saving human life at dangerous places. A robot, which could be controlled from a distant location, can solve these problems. This could be via communication waves or networking methods. Also controlling should be in real time and smooth so that it can actuate on every minor signal in an effective way. This paper discusses a method to control a robot over the network from a distant location. The robot was controlled by hand gestures which were captured by the live camera. A DSP board TMS320DM642EVM was used to implement image pre-processing and fastening the whole system. PCA was used for gesture classification and robot actuation was done according to predefined procedures. Classification information was sent over the network in the experiment. This method is robust and could be used to control any kind of robot over distance.

## 1. Introduction

There have been many robotic applications where a smooth and real time control is needed according to human requirements. Sometimes it is very difficult to transfer input to the robot using the keyboard as the human need to convert his input from ideas to keyboard values. It is tricky in the case of robot movement in different directions. Robot control through hand gesture is a developing area where a human can ask the robot to do something just by hand movements and no need to enter numeric values using the keyboard. Also, the input information could be transferred over the network to control a robot remotely. This can be used to control any type of robot for different actuation [1]. It can be used to control the robotic arm remotely where the human presence is not conducive for his/her safety or to control a TV or PC monitor in order to replace the infrared remote [2].

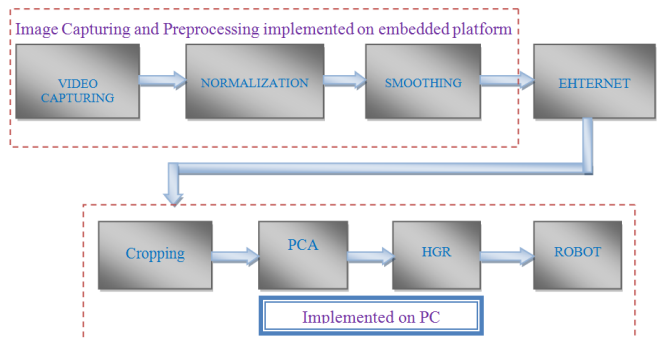

Fig 1. Block diagram for System

This paper demonstrates a system for controlling the robot by just showing hand movements in front of a camera. High-resolution camera connected to a Digital Signal Processing (DSP) based embedded board, which captures the image in real time and followed by the image pre-processing. In previous work, the algorithmic computations were performed on PC or some control device [3]. The pre-processing takes maximum time in any image processing algorithm implementation [4-5].

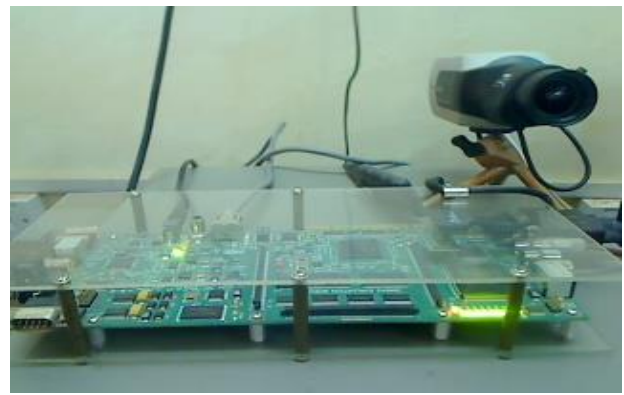

Fig 2. DSP processor (DM642EVM) with high-resolution camera

This work implements algorithms on a DSP processor TMS320DM642EVM to determine actuation using various hand gestures without touching the surface of the screen. This provides many fold increase in processing speed over PC implementations. Robust real-time performance is thus made possible and this increases the scope of applications to include ones where high frame rates are necessary.The block diagram of the system is shown in Fig. 1 while the DSP board setup is shown in Fig. 2.

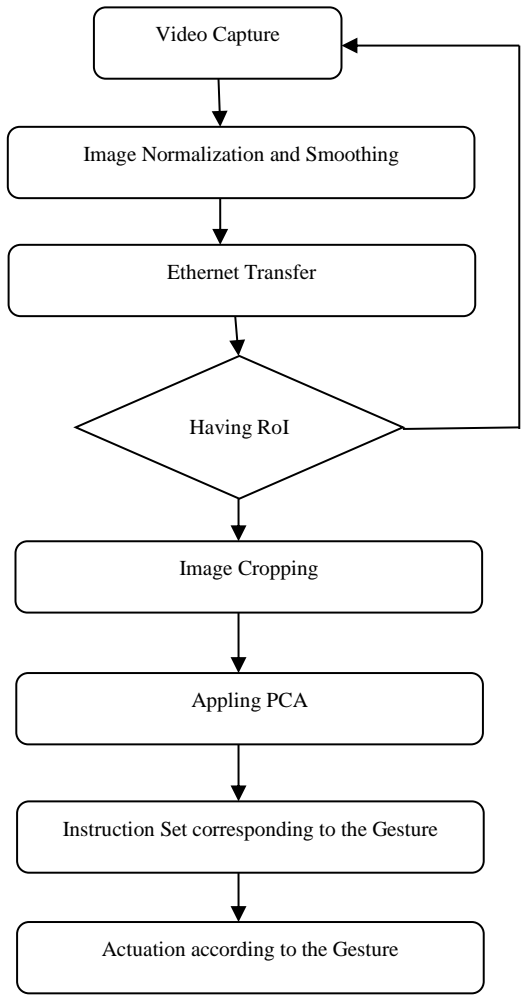

Fig 3. Process description

## 2.    Background

The Idea of tell-presence, as well as its past and future, is well described by Tachi [6]. He said that Telexistanece technology can allow for the highly realistic sensation in distant places without actual being presented there. He discussed many implementations of robots, made for teleoperation in the specific field.

Ma [7] presented a haptic glove with teleoperation experiments. This was capturing each finger's movement for operations. Li [8] showed virtual robot teleoperation which was based on hand gesture recognition. Adaptive Neuro-Fuzzy Inference Systems and SVM were used for classification.Wen [9] presented a hand gesture based robotic surgery based on augmented reality. Hand gesture signals were used to implements specified RF needle insertion plans.

Greenfield [10] discusses touchless hand gesture device controller with an image sensor. This method was based on hand motion and optical pattern recognition. DU [11] presented hand gesture based controlling of the robot manipulator. He used Kalman and particle filters to track hands. Although it was a markerless method but used leap motion to detect hand positions. Samsung has also built a TV which could be controlled using hand gesture from a certain distance [12].

Our presented method does not use any sensor or marking scheme. It is based on Natural Computing concept where anyone can use the system irrespective of skin color and hand size. The robot used in this experiment would also provide its surroundings visual so that controller can see what robot can do at that location. Principle component analysis (PCA) was used for classification. The process has been shown in Fig 3.

## 3.    Implementation

The developed system is using live video stream for gesture recognition. It captures frames from the live video in fixed time interval. The proposed technique to actuate robotic system is divided into four subparts:

- Gesture extraction from video stream
- Extract region of interest from a frame
- Determine gesture by matching patterns
- Determine actuate instruction and robot actuation

The working of the PC implementation of the system is explained in Fig 6.These techniques were implemented in Embedded Matlab, Simulink and Code Composer Studio (CCS). An XDS560 PCI JTAG Emulator was fitted in a PCI-slot to supports high-speed RTDX on the enabled processor for data transfer over 2 MB/second.

### A.  Gesture extraction from video stream

The video was captured by a high-resolution camera and image pre-processing techniques were applied on it frame by frame [13]. Two light sources were required to overcome shades, one below the camera and other on the top of the camera. The camera can be NTSC or PAL. Each frame was captured in each 1/5 second from the liver video stream. The preprocessing techniques such as smoothing and normalization reside on DSP processor DM642EVM. Fig. 4 shows these two functions being executed on the DSP board. More details about pre-processing on DSP board are discussed in [14-15].

This board brings a big reduction in execution time for the whole algorithm. After these steps, the image frame was transferred to PC using UDP target to HOST Ethernet communication as shown in Fig 5. Fig 6 shows the data being sent from PC to HOST. The remaining section of the system such as Cropping and PCA were running on PC.

Each gesture frame is analyzed to detect motion of the hand and to find where there was no or little motion. After applying global thresholding, each frame was transformed into two color image and then it was compared with two recently captured frames to find the changes in their pixel values. The difference between pixel values is counted with both frames and summed to measure total points of difference. Since the two color image has only values either 1 or 0, so XOR function can help in determining mismatched locations. There are three matrixes called *frame1, frame2, and frame3,* containing frames captured in three continuous time slot respectively then:

$$fr1 = frame1\ XOR\ frame3$$
$$fr2 = frame2\ XOR\ frame3$$
$$mismatch\_matrix = fr1\ OR\ fr2$$

A full mismatch is equal to the total number of 1s in matrix *mismatch_matrix*. If the point of a mismatch for any frame was detected less than 1% of a total number of pixels in that frame, it was a motionless frame (i.e. the frame which was supposed to be contained complete gesture). This frame would be selected for further operations of the system.

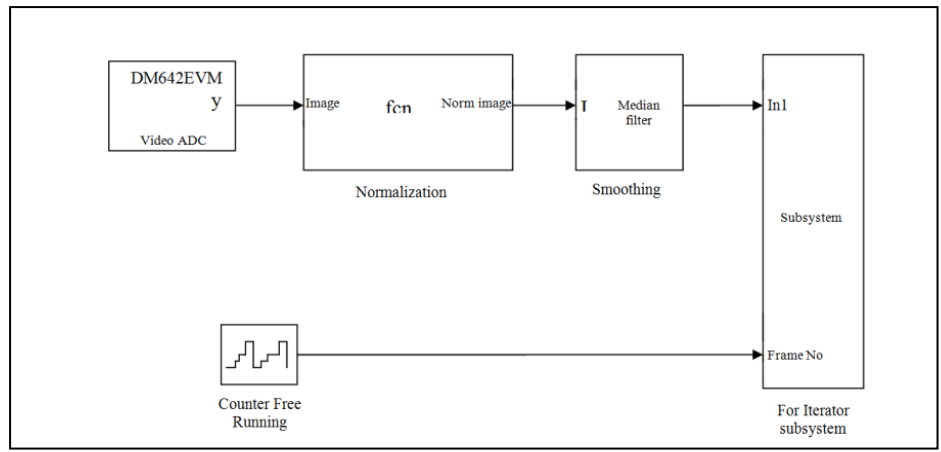

Fig 4. Image pre-processing on DSP board

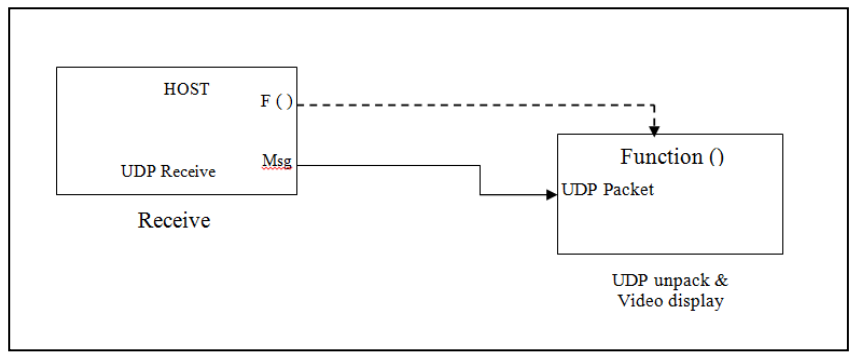

Fig 5. UDP Target to host Communication

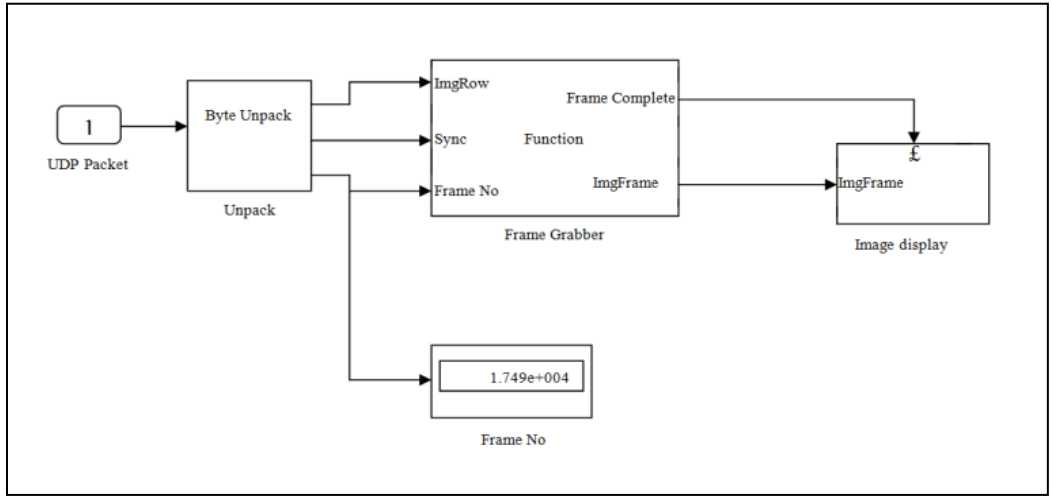

Fig 6. Data Transfer to Host

### B.  Extract ROI from selected frame

The frame may contain some extra information other than hand gesture, which was not requited. Therefore before proceeding further, these extra parts were removed. For cropping, selected frame was converted into a binary image using global thresholding. Then extra black rows and columns are determined from where the object of interest starts appearing. This is done by looking from each side of the two color image and moving forward till white pixels are more than offset value. The results show that offset value set to 1% of the total width gives a better result for overcoming noise. If size of the region of interest image is *m*n* then:

$$Hor\_offset = m/100$$
$$Ver\_offset = n/100$$

*Where Min_col and Max_col* represent minimum and maximum values of column no. where total number of white pixels are more than *Hor_offset*.
The *Min_row and Max_row represent* minimum and maximum values of row no. where total number of white pixels are more than *Ver_offset*.

After this, remove those parts of a hand which is not useful in gesture presentation in RoI i.e. removal of the wrist, a section of arm etc. Because theses extra information can give unwanted result due to limitations of the gesture database. So part of hand before wrist is to be cropped out. It has been shown by statistical analysis of hand shape that either it is palm or fist, width is going to be the lowest at the wrist and highest at the middle of the palm. Therefore the unwanted part of the human hand could be cropped out from the image by determining the location in the vertical histogram. The method could be applied as:

*Global Maxima = column no. where the height of histogram was highest*
*Cropping point = column no. where height of histogram was lowest*

Fig 9 shows a frame containing hand gesture before and after cropping.

### C.  Determining gesture and Pattern matching

Finally cropped gesture images were resized to matched database image sizes. Gesture database was created and stored in the binary form of image size 60x80. PCA is faster when compared to ANN which requires a long time in the training database and as well as high computation power. Also, different components in a human hand are big enough when compared to noise. PCA is was used for pattern matching [16].

### D.  Control instruction generation for Robot

The methods for each required gesture are written and stored in a database for controlling a robotic hand. PUMA robotic model was used for testing robotic control. Six different type of hand gestures were used for designing specific moves of the robot in precise angle and direction as shown in Fig 8. Movement commands have redesigned a module in the language understood by robot where it could actuate corresponding to each meaningful hand gesture.

When a gesture was presented in front of a camera, it was processed and the matched with stored gesture from the database. Action information corresponding to that gesture would be processed and passed to the robot for execution of stored commands. Immediately the robot performed action accordingly. The robot continued to do action according to gestures, shown by the user on camera. When the robot was moving, visual information of its surrounding was sent wirelessly to the user, as this robot was having a camera. Hand gesture processing, robot movement, and robot view are shown in Fig 7 respectively. It helped the user in obstacle avoidance and makes robot movement accordingly.

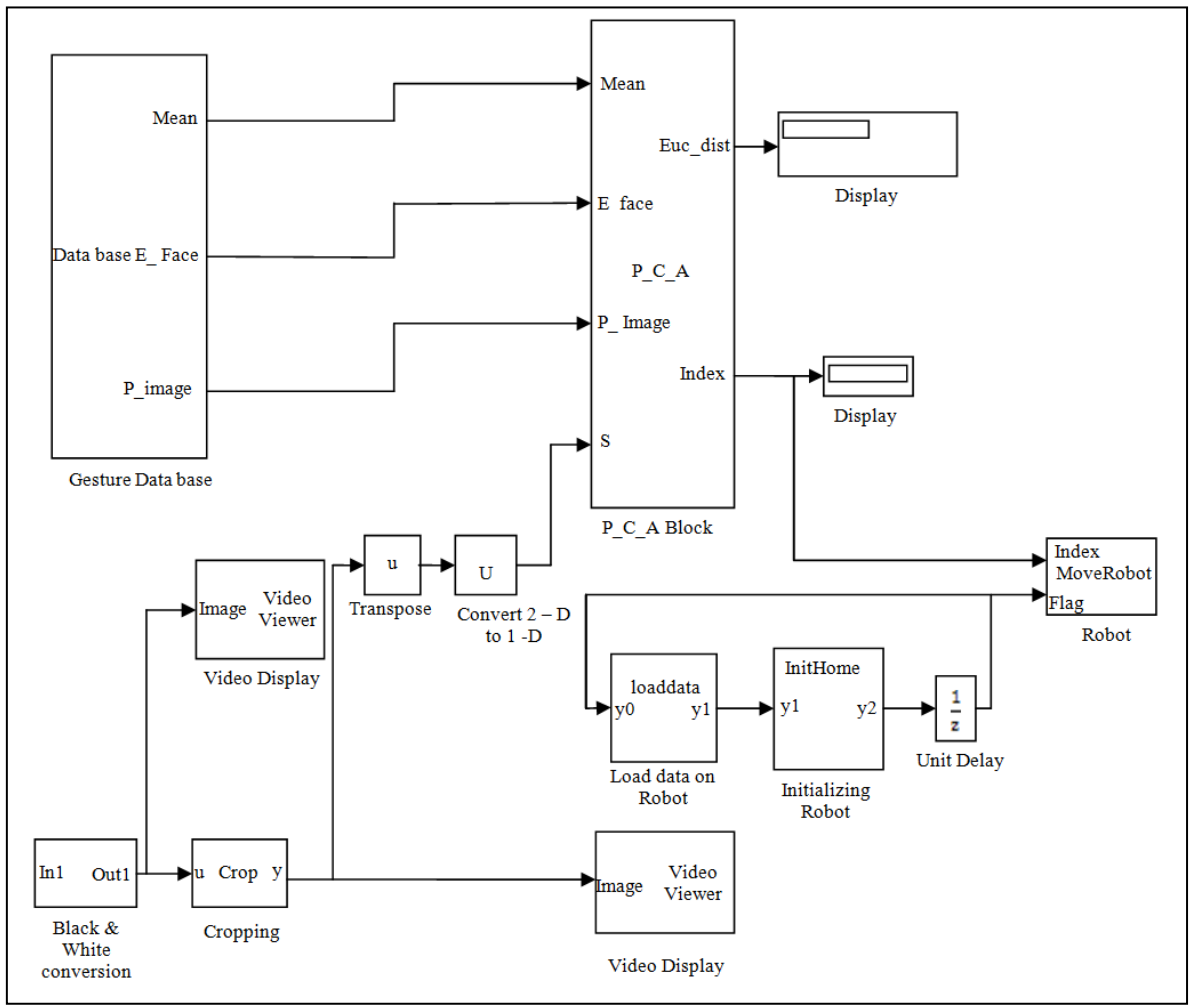

Fig 7.Implementation blocks diagram

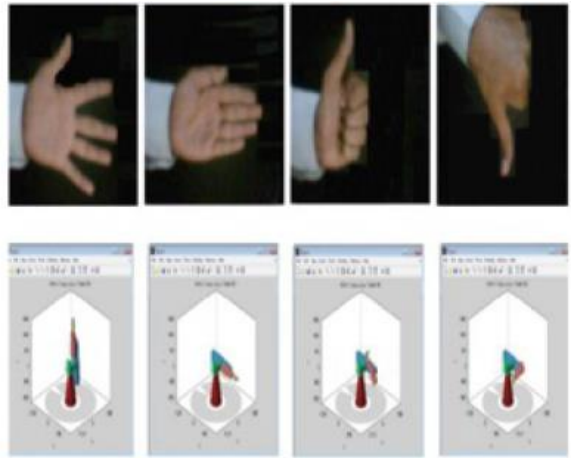

Fig 8.Different hand gestures

## 4.    Results and Conclusions

The need of telexistence in real-time inspired us to implement this system. A robot can be controlled over the network using hand gestures. The robot will move as per the gesture and would do movement and manipulation as per instruction. The proposed technique was tested in the environment which is shown in Fig 6. Database of gestures was stored into the binary format of 60x80 pixels so it was taking less time and memory during pattern recognition. Due to cropped image of gestures, the system becomes more effective as the one image is sufficient for one type of gesture presentation. So we need not to store more than one image for same gesture at different positions of the image. Also, we don't have to worry about positions of hand in front of the camera.

Experimental results show that system detects hand gesture when user stop moving hand for one second. The accuracy of the system is 95%. This method can be applied to any type of robot as Robot instructions were mapped on hand gesture signals. Currently, only ten gestures are implemented for experimental purpose, which could be extended as per requirement.

The system was implemented on a DSP board which brings its response time within real-time constraints. The robot is also providing visual information which could be used for different purposes including surveillance and object manipulation. In future, we would like to build the whole system-on-chip, so that it would be much smaller and low power consuming for better performance.

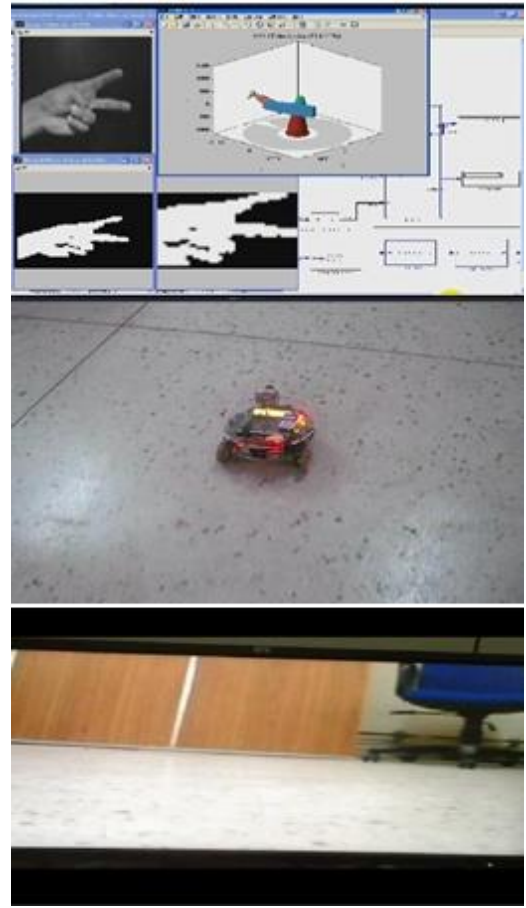

Fig 9. Robot movement and its vision

## Acknowledgement

This work has been carried out at CSIR-Central Electronics Engineering Research Institute, Pilani India. Authors are very thankful to the Director, CSIR CEERI for this active encouragement and continuous support.